\documentclass{acm_proc_article-sp}
\usepackage{gastex}
  \usepackage[draft,inline,nomargin]{fixme}
 \usepackage{color}
 \usepackage[alwaysadjust]{paralist}

\newenvironment{example}%
   {\vspace{-2ex}\begin{trivlist}\item[]\textsc{Example 1.}\itshape\ }%
   {\end{trivlist}}
\newenvironment{examplecont}%
   {\vspace{-2ex}\begin{trivlist}\item[]\textsc{Example 1 (cont.).}\itshape\ }%
   {\end{trivlist}}
\newcommand{\mydot}{\ensuremath{\!\cdot\!}} 
\newcommand{\chv}{\ensuremath{\mathbf{u}}} 
\newcommand{\bfx}{\ensuremath{\mathbf{x}}} 
\newcommand{\bfp}{\ensuremath{\mathbf{p}}} 
\newcommand{\bfk}{\ensuremath{\mathbf{k}}} 
\newcommand{\bfnull}{\ensuremath{\mathbf{0}}} 
 
\newcommand{\bfr}{\ensuremath{\mathbf{r}}} 
\newcommand{\bfy}{\ensuremath{\mathbf{y}}} 
\newcommand{\bfX}{\ensuremath{\mathbf{X}}} 
\newcommand{\bfV}{\ensuremath{\mathbf{V}}}
\newcommand{\bfv}{\ensuremath{\mathbf{v}}}
\newcommand{\bfw}{\ensuremath{\mathbf{w}}}
\newcommand{\bfW}{\ensuremath{\mathbf{W}}}  
\newcommand{\bfZ}{\ensuremath{\mathbf{Z}}}
\newcommand{\bfe}{\ensuremath{\mathbf{e}}} 
\newcommand{\bfz}{\ensuremath{\mathbf{z}}} 
\newcommand{\pr}[1]{\ensuremath{\mathit{Pr}\!\left(#1\right)}} 
 \newcommand{\sig}{\ensuremath{\mathit{Sig}}} 
 \newcommand{\coeff}{\ensuremath{\mathit{coeff}}} 
 \newcommand{\prob}{\ensuremath{\mathit{prob}}} 
 \newcommand{\field}{\ensuremath{\mathit{field}}} 
 \renewcommand{\mid}{\,|\,} 
 \newcommand{\dna}{\ensuremath{\mathit{DNA}}}

\title{Hybrid Numerical Solution of the\\ Chemical Master Equation}
 \numberofauthors{4}  
\author{
 \alignauthor
\begin{tabular}{l}\hspace*{-25.5ex}Thomas A. Henzinger\end{tabular}
\\[0.5ex]
       \affaddr{\hspace*{-25ex}\begin{tabular}{l}\begin{normalsize}Institute of Science and Technology, Austria                                                                                             \end{normalsize}\\
\end{tabular}}
      \alignauthor
      \begin{tabular}{l}\hspace*{-5.5ex}Maria Mateescu\end{tabular}\\[0.5ex]
       \affaddr{\begin{tabular}{l}\begin{normalsize}\,Ecole Polytechnique F\'{e}d\'{e}rale de Lausanne, Switzerland \end{normalsize} \\
\end{tabular}}
              \and
\alignauthor
\begin{tabular}{l}\hspace*{-25.5ex}Linar Mikeev\end{tabular}\\[0.5ex]
     \affaddr{\hspace*{-35.5ex}\begin{tabular}{l}\begin{normalsize}Saarland University, Germany                                                                           \end{normalsize}\\
\end{tabular}}
 \alignauthor
      \begin{tabular}{l}\hspace*{-9ex}Verena Wolf\end{tabular}
\\[0.5ex]
       \affaddr{\begin{tabular}{l}\begin{normalsize}Saarland University, Germany                                                              \end{normalsize}\\
\end{tabular}}
  }
 
\begin{document}
         
\maketitle
 
\begin{abstract}
 
We present a numerical approximation technique for the analysis of
continuous-time Markov chains that   describe 
networks of biochemical reactions and play
an important role in the stochastic modeling of biological systems.
Our approach is based on the construction of a stochastic 
hybrid model in which certain discrete random variables of the original
Markov chain are approximated by continuous
deterministic variables. 
We compute the solution of the   stochastic
hybrid model using a numerical algorithm that 
discretizes time and in each step 
performs a mutual update of the transient probability 
distribution of the discrete stochastic variables and 
the values of the continuous deterministic variables.
We implemented the algorithm and we demonstrate its usefulness and
efficiency on several case studies from systems biology.
\end{abstract}
%
%


\section{Introduction}
A common dynamical model in systems biology is a system of 
ordinary differential equations (ODEs) that describes 
the time evolution of the concentrations of certain proteins in a biological compartment.
This macroscopic model is based on the theory of chemical kinetics and assumes that the 
concentrations of chemical species in a well-stirred system change deterministically and 
continuously in time. 
It provides an appropriate description of a chemically reacting
 system as long as the numbers of molecules 
of the chemical species are large. 
However, in living cells the chemical populations can be low 
(e.g. a single DNA molecule, tens or a few hundreds of RNA or protein molecules). 
In this case the underlying assumptions of the ODE
approach are violated and  a more detailed model is necessary, which takes into account the inherently discrete and 
stochastic nature of chemical 
reactions~\cite{mcadams-arkin99,RaoWolf02,fedoroff-fontana,paulsson,swain}.
The theory of stochastic chemical kinetics provides an appropriate description by means of 
a discrete-state Markov process, that is, a continuous-time Markov chain (CTMC)
that represents the chemical populations as random variables~\cite{gillespie77,gillespie92}.
If $n$ is the number of different types of molecules, then we describe the 
state of the system at a certain time instant
by an $n$-dimensional random vector whose $i$-th entry represents
the number of molecules of type $i$.  
In the thermodynamic limit (when the number of molecules and the volume of the system 
approach infinity) the Markov model and the macroscopic ODE description
   are equal~\cite{kur81}.
   Therefore, the ODE approach
   can be used to approximate the  CTMC only if all populations are large.

The evolution of the CTMC is given by a system of linear ordinary differential 
equations, known as the \emph{chemical master equation} (CME).
A single equation in the CME describes the time derivative of the probability  
of a certain state at all times $t\ge 0$. Thus,  
the solution of the CME is the probability distribution over all  
states of the CTMC at a particular time $t$, 
that is, the transient state probabilities at time $t$.
The solution of the CME can then be used  to derive measures of interest
such as the distribution of
switching delays~\cite{mcadams-arkin97}, the distribution of the time of DNA
replication initiation at different origins~\cite{Patel2006}, or the
distribution of gene expression products~\cite{distr}.  
Moreover, many parameter estimation methods require the computation 
of the posterior distribution because means and variances do not provide enough
information to calibrate parameters~\cite{Wilkinsonpaper}.

The more detailed description of chemical reactions using a 
CTMC comes    at a price of significantly  increased computational complexity 
because the underlying state space is usually very large or even infinite.
Therefore, Monte Carlo simulation is  in widespread use, because  it allows
to generate random trajectories of the model  while requiring only little memory. 
Estimates of the measures of interest can be derived once the number 
of trajectories is large enough to achieve the desired statistical 
accuracy.
However, the main drawback of simulative solution techniques 
is that   a large number of trajectories is necessary to obtain 
reliable results.
For instance, in order to halve the confidence 
interval of an estimate, four times more trajectories have to be generated.
Consequently, often stochastic simulation is only feasible with a very
  low level of confidence in the accuracy of the results.

Recently, efficient numerical algorithms have been developed  to compute an 
approximation of the 
CME~\cite{sliding,Munsky06,krylovSidje,CMSB09,HIBI09,sparsegrid,Inexact,Engblom2009208,JahnkeWavelet}.
Many of them are based on the idea of restricting the analysis of the model during a certain time interval
 to a subset of states that have ``significant'' probability.
While some of these methods rely on an a priori estimation of the geometric bounds of
the significant subset~\cite{sliding,Munsky06,krylovSidje}, others are based on a conversion
to discrete time and they decide dynamically which states to consider  at a certain time step~\cite{CMSB09,HIBI09,Inexact}.

If the system under consideration contains large populations, then the 
numerical algorithms mentioned above perform poorly. 
The reason is that the random variables that represent large populations
have a large variance.  
Thus, a large number of states have a significant probability, which
renders the numerical approximation of the distribution computationally expensive or infeasible.

In this paper we use a stochastic hybrid approach to efficiently 
approximate the solution of systems containing both small and large populations.
More precisely, we maintain the discrete stochastic representation for small 
populations, but at the same time we exploit the small relative variance of 
large populations and represent them by continuous deterministic
variables.
Since population sizes change over time we decide dynamically ("on-the-fly")
whether we represent a population by a continuous deterministic variable
or keep the discrete stochastic representation. 
Our criterion for changing   from a discrete to a continuous 
treatment of a variable  and vice versa is based on a population threshold.

For the solution of the stochastic hybrid model, we propose a numerical 
approximation method that discretizes time and performs a mutual update 
of the distributions of the discrete stochastic variables and the values of the
continuous deterministic variables. 
Hence, we   compute the solution  of a CME with a reduced dimension
as well as the solution of a system of (non-linear) ordinary differential equations.
The former describes the distribution of the discrete stochastic variables and the
latter the values of the continuous deterministic variables, and 
  the two descriptions depend on each other. Assume, for instance, 
that a system has two chemical species.
The two population sizes at time $t$ are represented by the random 
variables $X(t)$ and $Y(t)$, where $X(t)$ is large and $Y(t)$ is small.
Then, we consider for $Y(t)$ all events $Y(t)=y$ that have significant 
probability, i.e., $\pr{Y(t)=y}$ is greater than a certain threshold. 
For $X(t)$ we consider the conditional expectations $E[X(t)\mid Y(t)=y]$
and assume that they change continuously and deterministically in time.
We iterate over small time steps $h>0$ and, given the distribution for $Y(t)$ and 
the values $E[X(t)\mid Y(t)=y]$, we compute the distribution of  $Y(t+h)$
and the values $E[X(t+h)\mid Y(t+h)=y]$. Again, we restrict our computation 
to those  values of $y$ that have significant probability.

To demonstrate the effectiveness of our approach, 
we have implemented the 
algorithm and applied it successfully to several examples from systems 
biology.
 Our most complex example has 6 different chemical species
 and 10 reactions. 
We compare our results with our earlier 
purely discrete stochastic approach and 
with the purely continuous deterministic approach in terms of
running times  and accuracy.

{\bf Related Work.}
  Different hybrid approaches have been proposed in the 
  literature~\cite{haseltine-rawlings,SalisKaznessis,Puchalka20041357}.
  As opposed to our approach, they focus on Monte Carlo simulation and 
  consider the problem of multiple time scales. 
  They do not use deterministic variables
  but try to reduce the   computational complexity of generating a trajectory
  of the model by approximating the number of reactions during a certain 
  time step.
   The closest work to ours is the hybrid approach proposed by
  Hellander and L\"otstedt~\cite{Hellander08}. 
  They approximate   large populations by normally distributed random variables 
    with a small variance and use Monte Carlo simulation to 
  statistically  estimate the probability 
  distribution of the remaining populations with small sizes.
They consider a single ODE to approximate the expected sizes of the large populations.
    As opposed to that, here we consider a set of ODEs to approximate the expected sizes 
of the large populations \emph{conditioned} on the small populations. 
This allows us to track the dependencies between the different populations more 
acurately. Moreover, instead of a statistical estimation of probabilities, we provide a  
    direct numerical method to solve the stochastic hybrid model.  
    The direct numerical method that we use for the computation of the 
    probability distributions of the stochastic variables  has shown to be
    superior to Monte Carlo simulation~\cite{CMSB09}.
    Another difference is that the method in~\cite{Hellander08} does not allow 
    a dynamic switching between stochastic and deterministic treatment
    of variables. 
    
    Finally, our approach  is related to the stochastic hybrid models 
      considered in~\cite{Lygeros,Lygeros2} 
     and to  fluid stochastic Petri nets~\cite{Horton}.
     These approaches differ from our approach in that they use probability
     distributions for the different values a continuous variable can take.
    In our setting, at a fixed point in time we only consider the conditional expectations
of the continuous variables, which is based on the assumption that the 
    respective populations are large and their relative variance is small. 
    This allows us to provide an efficient numerical approximation algorithm 
    that can be applied to systems with large state spaces. 
    The stochastic hybrid models in~\cite{Lygeros,Lygeros2,Horton} cannot be 
    solved numerically except in the case of small state spaces.

 \section{Discrete-state Stochastic\\ Model}\label{sec:cme}
 According to Gillespie's theory of stochastic chemical kinetics, 
 a well-stirred mixture of $n$ molecular species in a volume 
 with fixed size and fixed temperature can be represented as 
 a continuous-time Markov chain 
 $\left(\bfX(t),t\ge 0\right)$~\cite{gillespie77}. 
 The random vector $\bfX(t)=\left(X_1(t),\ldots,X_n(t) \right)$ 
 describes the chemical populations at time $t$, i.e., $X_i(t)$ is 
 the   number of molecules of type $i\in\{1,\ldots,n\}$.
Thus, the state space of $\bfX$ is $\mathbb Z^n_+=\{0,1,\ldots\}^n$.
The state changes of $\bfX$ are triggered by the occurrences of 
chemical reactions, which come in $m$ different types.  
For $j\in\{1,\ldots,m\}$ let $\chv_j\in\mathbb Z^n$ be the 
\emph{change vector} of the $j$-th reaction type, that is, 
$\chv_j=\chv_j^-+\chv_j^+$ where $\chv_j^-$ contains only 
non-positive entries that specify how many molecules of each
species are consumed (\emph{reactants}) if an instance of the 
reaction occurs and vector  $\chv_j^+$ contains only non-negative
 entries that specify how many molecules of each
species are produced (\emph{products}). 
Thus, if $\bfX(t)=\bfx$ for some $\bfx\in\mathbb Z^n_+$ with 
$\bfx+\chv_j^-$ being non-negative, then $\bfX(t+dt)=\bfx+\chv_j$
is the state of the system after the occurrence of the $j$-th 
reaction within the infinitesimal time interval $[t,t+dt)$.
 
As rigorously  derived by Gillespie~\cite{gillespie92}, each reaction 
type has an associated \emph{propensity function}, denoted by 
$\alpha_1,\ldots,\alpha_m$, which is such that 
$\alpha_j(\bfx)\cdot dt$ is the probability that, given $\bfX(t)=\bfx$, 
one instance of the $j$-th reaction occurs within $[t,t+dt)$. 
The value $\alpha_j(\bfx)$ is proportional to the number of distinct
reactant combinations in state $\bfx$. More precisely, if 
$\bfx=(x_1,\ldots,x_n)$ is a state for which $\bfx+\chv_j^-$ 
is nonnegative then
\begin{equation}\label{eq:propf}
\alpha_j(\bfx)=\left\{\!\!\begin{array}{l@{\,}c@{\,}l}
c_j & \mbox{ if } & \chv_j^-=(0,\ldots,0),\\
c_j \cdot x_i& \mbox{ if } & \chv_j^-=-\bfe_i,\\
c_j \cdot x_i\cdot x_\ell& \mbox{ if } & \chv_j^-=-\bfe_i-\bfe_\ell,\\
c_j \cdot {x_i\choose 2}=c_j \!\cdot\! \frac{x_i\cdot (x_i-1)}{2}& \mbox{ if } & \chv_j^-=-2\cdot \bfe_i,\\
\end{array}\right.
\end{equation}
where $ i\neq \ell$, $c_j>0$ is a constant, and 
$\bfe_i$ is the vector with the $i$-th entry $1$ and all other
entries $0$. 
We set $\alpha_j(\bfx)=0$ whenever the vector 
$\bfx+\chv_j^-$ contains negative entries, that is, 
when not enough reactant molecules are available.
The constant $c_j$ refers to the probability that a randomly selected 
pair of reactants collides and undergoes the $j$-th chemical reaction. 
Thus, if $N$ is the volume (in liters) times Avogadro's number, then $c_j$ 
\setlength{\pltopsep}{-4pt}
\begin{compactitem} 
\item scales inversely with $N$ in the case 
of two reactants, 
\item is independent of $N$ in the case of a 
single reactant, 
\item  is proportional to $N$ in the case of no reactants.
\end{compactitem} 
Since reactions of higher order (requiring 
more than two reactants) are usually the result of several successive lower 
order reactions, we do not consider the case of more than two reactants.
\begin{example}
We consider a gene regulatory
network, called the exclusive switch~\cite{exswitch}.
It consists of two genes with a common promotor region.
Each of the  two gene products $P_1$ and $P_2$ inhibits the 
expression of the other product if a molecule is bound to the 
promotor region. More precisely, 
if the promotor region is free, molecules of  both types 
$P_1$ and $P_2$ are produced.
If a molecule of type $P_1$ ($P_2$) is bound to the promotor 
region, only molecules of    type $P_1$ ($P_2$) are produced, respectively.
 We illustrate the system in Fig.~\ref{fig:switch}.
 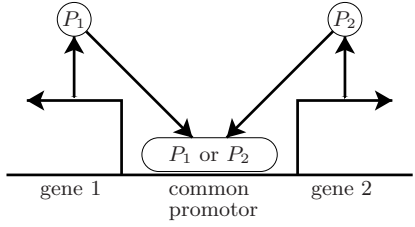
\begin{figure}[t]
\begin{center}
\scalebox{0.9}{ \begin{picture}(60,35)
  \gasset{Nw=5,Nh=5,Nmr=5}
\node(p1)(10,30){$P_1$}
\node(p2)(50,30){$P_2$}
\node[Nw=20](prom)(30,10){$P_1$ or $P_2$}
\drawline[AHnb=0,linewidth=0.4](0,7)(60,7)
\drawline[AHnb=1,linewidth=0.4,AHangle=30,AHLength=3,AHlength=2](17,7)(17,18)(3,18)
\drawline[AHnb=1,linewidth=0.4,AHangle=30,AHLength=3,AHlength=2](43,7)(43,18)(57,18)
  \gasset{Nw=0,Nh=0,Nmr=0,Nframe=y}
\node(pp1)(10,18.5){}
\node(pp2)(50,18.5){}
\drawedge[linewidth=0.4,AHangle=30,AHLength=3,AHlength=2](pp1,p1){}
\drawedge[linewidth=0.4,AHangle=30,AHLength=3,AHlength=2](pp2,p2){}
\drawedge[linewidth=0.4,AHangle=30,AHLength=3,AHlength=2](p1,prom){}
\drawedge[linewidth=0.4,AHangle=30,AHLength=3,AHlength=2](p2,prom){}
\put(5,4){gene 1}
\put(45,4){gene 2}
\put(24,4){common}
\put(24,1){promotor}
 \end{picture}}
\vspace{-3ex}
\end{center}		\caption{Illustration of the exclusive switch in Ex.~1 (picture is adapted from~\cite{exswitch}).
The stochastic hybrid model  with only three discrete stochastic states and two differential 
equations per state.\label{fig:switch}}
 \end{figure}
The system has five chemical species of which two 
have an infinite range, namely $P_1$ and 
$P_2$. If $\bfx=(x_1,\ldots,x_5)$ is the current state,
then the first two entries represent the populations of $P_1$ and $P_2$,
respectively. The entry $x_3$ denotes the number 
of unbound DNA molecules which is either zero or one.
The entry $x_{4}$ ($x_{5}$) is one of a molecule of type $P_1$ ($P_2$)
is bound to the promotor region and zero otherwise.
The chemical reactions are as follows. Let $j\in\{1,2\}$.
 	\begin{compactitem}
     
		 \item We describe production of  $P_j$
by $\dna\to\dna +P_j$. Thus, $\chv_j=\bfe_j$ and $\alpha_j(\bfx)=c_j\cdot x_3$.
\item We describe degradation of  $P_j$
by $P_j\to\emptyset$ with $\chv_{j+2}=-\bfe_j$ and $\alpha_{j+2}(\bfx)=c_{j+2}\cdot x_j$.
\item We model the binding of $P_{j}$  to the promotor 
by $\dna+P_j\to \dna.P_j$ with 
 $\chv_{j+4} =-\bfe_j-\bfe_3+\bfe_{j+3}$ and 
$\alpha_{j+4}(x)=c_{j+4} \cdot x_j\cdot x_3$. 
\item For unbinding of $P_j$ we use
  $\dna.P_j \to \dna+P_j$ with 
 $\chv_{j+6}(x)=\bfe_j+\bfe_3-\bfe_{j+3}$ and 
$\alpha_{j+6}(x)=c_{j+6}\cdot x_{j+3}$.
 \item Finally, we have production of $P_j$ if a molecule
of type $P_j$ is bound to the promotor, i.e.,
  $\dna.P_j \to \dna.P_j+P_j$ with  $\chv_{j+8}(x)=\bfe_j$ and 
$\alpha_{j+8}(x)=c_{j+8}\cdot x_{j+3}$.
 \end{compactitem}
 Depending on the chosen parameters, the probability distribution of the exclusive 
switch is bistable, i.e. most of the probability mass concentrates on two distinct regions in 
the state space. In particular, if   binding to the promotor is likely, then these two regions 
correspond to the two configurations where either the production of $P_1$ or the production of
$P_2$ is inhibited.
\end{example}

{\bf The Chemical Master Equation.}
For $\bfx\in \mathbb Z^n_+$ and $t\ge 0$, let $p(\bfx,t)$ denote the 
probability that the current population vector is $\bfx$, i.e., $p(\bfx,t)=\pr{\bfX(t)=\bfx}$. 
Let $\bfp(t)$ be the row vector with entries $p(\bfx,t)$.
Given $\chv_1^-,\ldots, \chv_m^-$, $\chv_1^+,\ldots, \chv_m^+$, 
$\alpha_1,\ldots,\alpha_m$, and some initial distribution $\bfp(0)$, the  
Markov chain $\bfX$ is uniquely specified if the propensity functions
are of the form in Eq.~\eqref{eq:propf}.    
The evolution of $\bfX$  is given by the chemical master equation (CME), 
which equates the change $\frac{d}{dt} p(\bfx,t)$
of the probability in state $\bfx$ and the sum over all 
reactions of the ``inflow'' $\alpha_j(\bfx-\chv_j)\cdot p(\bfx-\chv_j,t)$ and 
``outflow'' $\alpha_j(\bfx)\cdot p(\bfx,t)$ of probability~\cite{kampen}.
Thus, 
\begin{equation}\label{eq:CME}
\frac{d}{dt} p(\bfx,t)=  \sum_{j=1}^m\big(\alpha_j(\bfx-\chv_j)\cdot p(\bfx-\chv_j,t)
-  \alpha_j(\bfx)\cdot p(\bfx,t)\big).
\end{equation}
 Since the CME is linear
  it can be written as $\frac{d}{dt} \bfp(t)=\bfp(t)\cdot Q$, where $Q$ is the generator matrix of $\bfX$
  with $Q(\bfx,\bfx+\chv_j)=\alpha_j(\bfx)$ and 
  $Q(\bfx,\bfx)=-\sum_{j=1}^m \alpha_j(\bfx)$.
  If $Q$ is bounded, then Eq~\eqref{eq:CME} has the general solution 
\begin{equation} 
 \label{eq:sol}
  {\mathbf p}(t)={\mathbf p}(0)\cdot e^{Qt},
 \end{equation}
where the matrix exponential 
is defined as $e^{Qt}=  \sum_{i=0}^{\infty}\frac{(Qt)^i}{i!}.$
If the state space is infinite, 
then we can only compute approximations of  ${\mathbf p}{(t)}$
 and even if $Q$ is finite, the size of the matrix $Q$ is often large 
because it grows exponentially with the number of state variables. 
Moreover, even if $Q$ is sparse, as it usually is because the number of reaction 
types is small compared to the number of states, 
standard numerical solution techniques 
for  systems of  first-order linear equations of the form of Eq.~\eqref{eq:CME},
such as uniformization~\cite{jensen},  approximations in the
Krylov subspace~\cite{Scientifique95transientsolutions}, 
 or numerical integration~\cite{stewart},
are infeasible.
The reason is that the number  of nonzero entries in $Q$ often 
exceeds the available 
 memory capacity  for systems of realistic size. 
 If  the populations of all species remain
 small (at most a few hundreds) 
 then the solution of the CME can be efficiently
 \emph{approximated} using projection methods~\cite{sliding,Munsky06,krylovSidje}
 or fast uniformization methods~\cite{CMSB09,HIBI09,Inexact}.
The idea of these methods is to avoid an exhaustive state space 
exploration and, depending on a certain time interval, restrict the analysis
of the system to a subset of states.

{\bf Fast Solution of the Discrete Stochastic Model.} 
Here, we present a method similar  to our previous work~\cite{HIBI09}
that efficiently approximates the solution of  the CME if the chemical 
populations remain small. 
We use it in Section~\ref{sec:stochhybrid} to solve the discrete part of the 
stochastic hybrid model.
 
The algorithm, called fast RK4, is based on the numerical integration 
of Eq.~\eqref{eq:CME} using an explicit fourth-order Runge-Kutta method.  
The main idea is to integrate only 
those differential equations in Eq.~\eqref{eq:CME} that correspond to states 
with ``significant probability''. This reduces the computational effort significantly
since in each iteration step only a comparatively small subset of states is 
considered. We dynamically decide 
which states to drop/add based on a fixed probability threshold $\delta>0$.
Due to the regular structure of the Markov model the approximation error of the 
algorithm remains small since probability mass is usually concentrated at certain
parts of the state space. The farther away a state is from such a ``significant set''
the smaller is its probability. Thus, the total error of the approximation 
remains small. 
Unless otherwise specified, in our experiments we fix $\delta$ to  $10^{-14}$.
This  
value that has been shown to lead to accurate approximations~\cite{HIBI09}.

The standard explicit fourth-order Runge-Kutta method applied to Eq.~\eqref{eq:CME}
yields the iteration step~\cite{stewart}
\begin{equation}\label{eq:fRK}
\bfp(t+h)=\bfp(t) + h\cdot (\bfk_1+2\cdot \bfk_2 +2 \cdot \bfk_3+\bfk_4)/6,
\end{equation}
   where $h>0$ is the time step of the method and the vectors $\bfk_1,\bfk_2,\bfk_3,\bfk_4$
   are given by
\begin{equation}\label{eq:ks}\hspace{0ex}
\begin{array}{l@{\quad\ }l}
\bfk_1= \bfp(t)\cdot Q, & \bfk_3 = (\bfp(t)+h\cdot \frac{\bfk_2}{2})\cdot Q,\\[1ex]
\bfk_2 = (\bfp(t)+h\cdot \frac{\bfk_1}{2})\cdot Q,&
\bfk_4 = (\bfp(t)+h\cdot \bfk_3)\cdot Q.\\[1ex]
\end{array}
\end{equation}
Note that the entries $k_1(\bfx),\ldots,k_4(\bfx)$  of state $\bfx$ in the vectors $\bfk_1,\ldots,\bfk_4$
are given by
\begin{equation}\label{eq:ksdetails}
\hspace{-1ex}\begin{array}{r@{\,}c@{\,\!}l}
k_1(\bfx)&=&  \sum\limits_{j=1}^m\big(\alpha_j(\bfx\!-\!\chv_j)\!\cdot\! p(\bfx\!-\!\chv_j,t)
 \!-\!  \alpha_j(\bfx)\!\cdot\! p(\bfx,t)\big),\\[1ex]
k_{i+1}(\bfx) &=& \sum\limits_{j=1}^m\big(\alpha_j(\bfx\!-\!\chv_j)\!\cdot\! (p(\bfx\!-\!\chv_j,t)\!+\!h\!\cdot\! k_i(\bfx\!-\!\chv_j)/2)\\ & &
 -  \alpha_j(\bfx)\!\cdot\! (p(\bfx,t)\!+\!h\cdot k_i(\bfx)/2)\big)\, \mbox{ for } i\in \{1,2\},\\[1ex]
 k_4(\bfx) &=& \sum\limits_{j=1}^m\big(\alpha_j(\bfx\!-\!\chv_j)\!\cdot\! (p(\bfx\!-\!\chv_j,t)\!+\!h\cdot k_3(\bfx\!-\!\chv_j))\\ & &
 -  \alpha_j(\bfx)\!\cdot\! (p(\bfx,t)\!+\!h\cdot k_3(\bfx))\big).
\end{array}\hspace{-1ex}
\end{equation}
 In order to avoid the explicit construction of $Q$
and in order to work with a dynamic set $\sig$ of significant states
that changes in each step, we use for  a state $\bfx$ a data structure
 with the following components:
\begin{compactitem} 
\item a field $\bfx.\prob$ for the current probability of $\bfx$, 
\item fields $\bfx.k_1,\ldots, \bfx.k_4$ for the  four terms in
the equation of state $\bfx$ in the system of Eq.~\eqref{eq:ks}, 
\item for all $j$ with $\bfx+\chv^-_j\ge \bfnull$ 
 a pointer to the successor state $\bfx+\chv_j$ 
as well as the rate $\alpha_j(\bfx)$. 
\end{compactitem} 
We start at time $t=0$ and initialize the set $\sig$ 
as the set of all states that have 
initially a probability greater than $\delta$, 
i.e. $\sig:=\{\bfx\mid p(\bfx,0)>\delta\}$.  
We perform a step of the iteration in Eq.~\eqref{eq:fRK} by traversing the 
set $\sig$ five times. In the first four rounds we compute
$\bfk_1,\ldots,\bfk_4$ and in the final round we accumulate the summands. 
While processing state $\bfx$ in round $i$, $i<5$, for each reaction $j$, we transfer probability 
mass from state $\bfx$ to its successor $\bfx+\chv_j$, by subtracting a term from
$k_i(\bfx)$ (see Eq.~\eqref{eq:ksdetails}) and adding the same term to  
$k_i(\bfx+\chv_j)$.
A single iteration step is illustrated in pseudocode in Table~\ref{alg:fRK4}. 
In line 20, we ensure 
that $\sig$ does not contain states with a probability less than $\delta$.
We choose the step size $h$  in line 1
as suggested in \cite{stewart}.
In line 2-15 we compute the values $k_1(\bfx),\ldots, k_4(\bfx)$
for all $\bfx\in\sig$ (see Eq.~\eqref{eq:ks}). The fifth round starts in line 16
and in line 17 the approximation of the probability $p(\bfx,t+h)$
is calculated. Note that the fields   $\bfx.k_1,\ldots, \bfx.k_4$ are initialized 
with zero.

\begin{table}[t] \renewcommand\arraystretch{1.3}
\centering
\begin{tabular}{|l|l|} 
\hline 
 \multicolumn{2}{|l|}{1  \hspace{0ex} choose step size $h$;} \\[-0.5ex] 
\multicolumn{2}{|l|}{2 \hspace{0ex} \textbf{for} $i=1,2,3,4$  \textbf{do} \emph{/\!/traverse $\sig$ four times} }
\\[-0.5ex] 
\multicolumn{2}{|l|}{3 \hspace{2ex} \emph{/\!/decide which fields from state 
data structure} }  \\[-0.5ex] 
\multicolumn{2}{|l|}{4 \hspace{2ex} \emph{/\!/are needed for $k_i$} }  \\[-0.5ex] 
 \multicolumn{2}{|l|}{5 \hspace{2ex} \textbf{switch} $i$}
\\[-0.5ex] 
\multicolumn{2}{|l|}{6 \hspace{4ex} \textbf{case} $i=1$: $\coeff:=1$; $\field:=\prob;$}\\[-0.5ex] 
\multicolumn{2}{|l|}{7 \hspace{4ex}  \textbf{case} $i\in\{2,3\}$: $\coeff:=h/2$; $\field:=k_{i-1};$}\\[-0.5ex] 
\multicolumn{2}{|l|}{8  \hspace{4ex}  \textbf{case} $i=4$: $\coeff:=h$; $\field:=k_{i-1};$}\\[-0.5ex] 
\multicolumn{2}{|l|}{9  \hspace{2ex}    $\bfx.k_i:=\bfx.k_1;$}\\[-0.5ex] 
\multicolumn{2}{|l|}{10  \hspace{1.1ex} \textbf{for all} $\bfx\in \sig$ \textbf{do}} \\[-0.5ex] 
\multicolumn{2}{|l|}{11  \hspace{3.1ex} \textbf{for} $j=1,\ldots,m$ with $\bfx+\chv_j\ge \bfnull$ \textbf{do}} \\[-0.5ex] 
\multicolumn{2}{|l|}{12  \hspace{5.1ex}$\bfx.k_i:=\bfx.k_i-\coeff\cdot \bfx.field\cdot 
\alpha_j(\bfx)$;} 
\\[-0.5ex] 
\multicolumn{2}{|l|}{13 \hspace{5.1ex}\textbf{if} $\bfx\!+\!\chv_j \not\in \sig$ \textbf{then}}
\\[-0.5ex] 
\multicolumn{2}{|l|}{14 \hspace{7.1ex}  $\sig:=\sig\cup\{\bfx\!+\!\chv_j\}$;}\\[-0.5ex] 
\multicolumn{2}{|l|}{15 
\hspace{5.1ex}$(\bfx\!+\!\chv_j).k_i:=(\bfx\!+\!\chv_j).k_i+
\coeff\mydot \bfx.field\mydot \alpha_j(\bfx);\!\!$}\\[-0.5ex] 
\multicolumn{2}{|l|}{16  \hspace{-0.5ex} \textbf{for all} $\bfx\in \sig$ \textbf{do}}
\\[-0.5ex] 
\multicolumn{2}{|l|}{17  \hspace{1.5ex}   $\bfx.\prob\!:=\!\bfx.\prob\!+\!
h\mydot (\bfx.k_1\!+\!2\mydot  \bfx.k_2\!+\!2\mydot  \bfx.k_3\!+\!  \bfx.k_4)/6;\!\!$ }
\\[-0.5ex] 
\multicolumn{2}{|l|}{18  \hspace{1.5ex}   $\bfx.k_1:=0;\bfx.k_2:=0;
\bfx.k_3:=0;  \bfx.k_4:=0;$ }\\[-0.5ex] 
\multicolumn{2}{|l|}{19  \hspace{1.5ex}  \textbf{if} $\bfx.\prob<\delta$ \textbf{then}}
\\[-0.5ex] 
\multicolumn{2}{|l|}{20  \hspace{3.5ex}  $\sig:=\sig\setminus\{\bfx\};$ }
 \\[-0.0ex] 
\hline
\end{tabular} 
\caption{A single iteration step of the fast RK4 algorithm, which
 approximates the solution of the 
CME.\label{alg:fRK4}}
\end{table}

 Clearly, for the solution of the CME the same ideas as above 
  can be used for many other numerical integration methods.
Here, we focus on the explicit RK4  method and do not 
consider more advanced numerical integration methods 
to keep our presentation simple. 
The focus of this paper is not on particular numerical methods
 to solve differential equations  but rather on general strategies
 for the approximate solution of the stochastic models that we consider.
 Moreover, we do not use uniformization methods as in previous work
 since uniformization is inefficient for very small time horizons. 
 But small time steps are necessary for the solution of the hybrid model
 in order to take into account the dependencies between the stochastic
 and the deterministic variables.

\section{Derivation of the Deterministic Limit}
\label{sec:ODE}
The numerical approximation presented in the previous section 
works well as long as only the main part of the probability mass is concentrated 
on a small subset of the state space.
If the system contains large populations then the 
probability mass distributes on a very large number of states
whereas the information content is rather low since we distinguish,
for instance, the cases of having $X_i(t)=10000$, $X_i(t)=10001$,
etc. In such cases no direct numerical approximation of the CME
is possible without resorting to Monte Carlo techniques or 
discarding the discreteness of the state space.
   If all populations are large the solution of $\bfX$
can be accurately 
approximated by considering the deterministic limit of $\bfX$.
Here, we shortly recall the basic steps for the derivation of the deterministic limit.
For a detailed discussion, we refer to Kurtz~\cite{kur81}. 

We first define a set of functions $\beta_j$ such that 
if $N$ is large (recall that $N$ is the volume times the 
Avogadro's number) then
the propensity functions  can be approximated as 
 $\alpha_j(\bfx)\approx N\cdot \beta_j(\bfz)$, where
   $\bfz=(z_1,\ldots,z_n)=\bfx\cdot N^{-1}$ corresponds to the vector of concentrations of chemical
species and belongs to $\mathbb R^n$.
Recall the dependencies of 
$c_j$ on the scaling factor $N$ as described at the 
beginning of Section~\ref{sec:cme}. 
For constants $k_j>0$ that are \emph{independent} of $N$,
\begin{compactitem} 
\item $c_j = k_j \cdot N$ in the case of no reactants,
\item $c_j = k_j$ in the case of a single reactant, 
\item $c_j = k_j/N$ in the case of two reactants.
\end{compactitem} 
From this, it follows that except for the case of bimolecular reactions,
we can construct the functions $\beta_j$ such that $\alpha_j(\bfx) = N \cdot \beta_j(\bfz)$.
$$
\beta_j(\bfz)=\frac{\alpha_j(\bfx)}{N} = \left\{\!\!\begin{array}{l@{\,}c@{\,}l}
\frac{c_j}{N}=k_j & \mbox{ if } & \chv_j^-=(0,\ldots,0),\\
c_j\mydot \frac{x_i}{N}=k_j \mydot z_i& \mbox{ if } & \chv_j^-=-\bfe_i,\\
c_j\mydot x_i\mydot \frac{x_\ell}{N} =
k_j \mydot z_i\mydot z_\ell& \mbox{ if } & \chv_j^-=-\bfe_i-\bfe_\ell,\\
\end{array}\right.
$$
where $i\neq \ell$.  
In the case of bimolecular reactins ($\chv_j^-=-2\cdot \bfe_i$),
we use the approximation   
$$\begin{array}{lcl}
N \mydot \beta_j(\bfz)&=&k_j \mydot N\mydot z_i^2=k_j \mydot x_i\mydot z_i=
(\frac{1}{2} c_j N)  \mydot x_i\mydot z_i\\
&=& \frac{1}{2} c_j \mydot x_i^2 \approx \frac{1}{2} c_j \mydot x_i (x_i -1)=\alpha_j(\bfx),
\end{array}$$
which is accurate if $x_i$ is large, 
 In order to derive the deterministic limit for the vector 
$\bfX(t)=(X_1(t),\ldots,X_n(t))$ that describes the chemical populations, we first write 
$\bfX(t)$ as 
$$
\bfX(t)=\bfX(0)+\sum_{j=1}^m \chv_j\cdot C_j(t),
$$
 where $\bfX(0)$ is the initial population vector and $C_j(t)$ denotes the number of occurrences of
the $j$-th reaction until time $t$. The process  $C_j(t)$ is a counting process with intensity 
$\alpha_j(\bfX(t))$ and  it can be regarded as a Poisson process whose time-dependent 
intensity changes according to the stochastic process $\bfX(t)$. 
Now, recall that a Poisson process $\tilde Y(t)$ with time-dependent 
 intensity $\lambda(t)$ can be transformed into a Poisson process 
 $Y(u)$ with constant intensity one, using 
the simple time transform $u=\int_{0}^t \lambda(s)ds$, that is, 
$Y(u)=Y(\int_{0}^t \lambda(s)ds)=\tilde Y(t)$. Similarly,
we can describe $C_j(t)$ as a Poisson process with intensity one, i.e.,
 $$C_j(t)=Y_j\left(\int_{0}^t \alpha_j(\bfX(s)) ds \right),$$
where $Y_j$ are independent Poisson processes with intensity one.
Hence, for $i\in\{1,\ldots,n\}$
\begin{equation}\label{eq:count}
 X_i(t)=X_i(0)+\sum_{j=1}^m u_{ji}\cdot Y_j\left(\int_{0}^t \alpha_j(\bfX(s)) ds \right),
\end{equation}
where $\chv_j=(u_{j1},\ldots,u_{jn})$.
The next step is to define $\bfZ(t)=\bfX(t)\cdot N^{-1}$, that is,
 $\bfZ(t)=(Z_1(t),\ldots,Z_n(t))$   contains the concentrations of 
the chemical species in moles per liter
at time $t$.  Thus,
\begin{equation}\label{eq:step1}
 Z_i(t) =  Z_i(0) +\sum_{j=1}^m  u_{ji}\cdot  N^{-1}\cdot
 Y_j\left(\int_{0}^t \alpha_j(\bfX(s)) ds \right),
\end{equation}
and using the fact that $\alpha_j(\bfx)\approx N\cdot \beta_j(\bfz)$
yields
\begin{equation}\label{eq:step2}
 Z_i(t)\approx Z_i(0)+\sum_{j=1}^m u_{ji}\cdot  N^{-1}\cdot
 Y_j\left({N\cdot\int_{0}^t  \beta_j(\bfZ(s)) ds} \right).
 \end{equation}
By the law of large numbers, the unit Poisson process 
$Y_j$ will approach $N\cdot u$ at time $N\cdot u$ for large $N\cdot u$. 
Thus, $Y_j(N\cdot u)\approx N\cdot u$ and hence,
\begin{equation}\label{eq:step3}
 Z_i(t)\approx Z_i(0)+\sum_{j=1}^m u_{ji}\cdot   
 {\int_{0}^t  \beta_j(\bfZ(s)) ds}.
\end{equation}
The right-hand side of the above integral equation is the solution $\bfz(t)$ of the system of ODEs
\begin{equation}\label{eq:step4}
\frac{d}{dt} \bfz(t) = \sum_{j=1}^m \chv_j \cdot \beta_j(\bfz(t)).
\end{equation}
As shown by Kurtz~\cite{kur81}, in the large volume limit,
where the volume and the number of molecules 
approach infinity (while the concentrations remain constant),
 $\bfZ(t)\to \bfz(t)$ in probability for finite times $t$.
 Note that the chemical concentrations $\bfz(t)$ evolve 
continuously and deterministically in time.  
This continuous deterministic approximation is reasonable if  
all species have a small relative variance and if they are only 
weakly correlated.
The reason is that only in this case the assumption that $Z_i(t)$
is deterministic is appropriate.
Note that for most models this is the case if the population
of species $i$ is large since this implies that $E[X_i(t)]$ is large whereas
the occurrence of chemical reactions results only in a 
marginal relative change of the value of $X_i(t)$.
\begin{examplecont}
The ODEs of the exclusive switch   are given 
by 
$$
\begin{array}{r@{\,}c@{\,}l}
 \frac{d}{dt} z_1(t) &=& k_1\cdot z_3(t) -  k_3\cdot z_1(t)
-k_5\cdot z_1(t)\cdot z_3(t)\\
& &+k_7\cdot z_4(t)+k_9\cdot z_4(t)\\[1ex]
 \frac{d}{dt} z_2(t) &=& k_2\cdot z_3(t) - k_4\cdot z_2(t)
-k_6\cdot z_2(t)\cdot z_3(t) \\
& & + k_8\cdot z_5(t)+k_{10}\cdot z_5(t)\\[1ex]
 \frac{d}{dt} z_3(t) &=& -k_5\cdot z_1(t)\cdot z_3(t)-k_6\cdot z_2(t)\cdot z_3(t)\\ &&
+k_7\cdot z_4(t)+k_8\cdot z_5(t)
 \\[1ex]
 \frac{d}{dt} z_4(t) &=& k_5\cdot z_1(t)\cdot z_3(t)-k_7\cdot z_4(t) \\[1ex]
 \frac{d}{dt} z_5(t) &=& k_6\cdot z_2(t)\cdot z_3(t)-k_8\cdot z_5(t) \\[1ex]
 \end{array}
$$
where $z_1(t),z_2(t),z_3(t),z_4(t),z_5(t)$ denote the 
respective chemical concentrations.
Moreover,   $c_j=N^{-1}\cdot k_j$ for $j\in\{5,6\}$ and 
$c_j=k_j$ for $j\not\in\{5,6\}$.
     \end{examplecont}
In~\cite{Kurtz2006}, Ball et al. scale only a subset of the populations in order 
to approximate the behavior of the system if certain populations are large 
and others are small. Additionally, they take into account the different speeds of 
the chemical reactions. For a selected number of examples, 
they give analytical expressions for the distributions in the limit,
i.e., when the scaling parameter approaches infinity. In the next section,
we will construct a stochastic hybrid model that is equivalent to the one considered in
\cite{Kurtz2006} if we scale the continuous components and consider the deterministic limit.

\section{Stochastic Hybrid Model}\label{sec:stochhybrid}

A straightforward consequence of the CME is that the time derivative of the 
populations' expectations 
are given by
\begin{equation}\label{eq:exp}
\textstyle\frac{d}{dt} E[\bfX(t)] = \sum_{j=1}^m \chv_j \cdot E\left[\alpha_j\left(\bfX(t)\right)\right].
\end{equation}
If all reactions of the system involve at most one reactant, Eq.~\eqref{eq:exp} 
can be simplified to
\begin{equation}\label{eq:recrateeq}
\textstyle \frac{d}{dt} E[\bfX(t)] = \sum_{j=1}^m \chv_j \cdot \alpha_j\left(E[\bfX(t)]\right).
\end{equation}
because the propensity functions $\alpha_j$ 
are linear in $\bfx$. 
But in the case of bimolecular reactions, we have either $\alpha_j(\bfx)=c_j\cdot x_i\cdot x_\ell$
for some $i,\ell$ with $i\neq \ell$ or $\alpha_j(\bfx)=c_j\cdot x_i\cdot (x_i-1)/2$ if the $j$-th reaction involves two reactants 
of type $i$. But this means that 
$$ \begin{array}{l}
E\left[\alpha_j\left(\bfX(t)\right)\right]=c_j \cdot E\left[X_i(t) \cdot X_\ell(t)\right] \quad\mbox{ or }
 \\[2ex]
E\left[\alpha_j\left(\bfX(t)\right)\right]=\frac{1}{2}\cdot c_j \cdot E\left[(X_i(t))^2 \right] - E\left[(X_i(t)) \right],
   \end{array}
$$
respectively. In both cases new differential equations are necessary to describe the unknown 
values of $E\left[X_i(t) \cdot X_\ell(t)\right]$ and $ E\left[(X_i(t))^2 \right]$. This problem repeats 
and leads to an infinite system of ODEs. 
As shown in the sequel,
we can, however, exploit Eq.~\eqref{eq:exp} to derive a stochastic hybrid model.

 Assume we have a system where certain species have 
a large population. 
In that case  we approximate them with  continuous deterministic
variables. The remaining variables are kept  discrete stochastic.
 This is done because it is usually infeasible or at least computationally very
costly to solve a purely stochastic model with high populations 
since in the respective dimensions the number of significant states is
large. 
Therefore, we propose to switch to a hybrid model where the stochastic part does not
contain large populations. In this way we can guarantee an efficient
 approximation of the solution.

Formally, we split $\bfX(t)$ into {small} populations $\bfV(t)$
and large populations $\bfW(t)$, i.e.  
$\bfX(t)=\left({\bfV(t)},{\bfW(t)}\right)$.
Let $\tilde n$ be the dimension of $\bfV(t)$  and $\hat n$
   the dimension of $\bfW(t)$, i.e. $n=\tilde n+\hat n$. 
Moreover, let $\tilde D$ and $\hat D$ be the set of indices 
that correspond to the populations in $\bfV$  and $\bfW$, respectively.
Thus, $\tilde D,\hat D\subseteq\{1,\ldots,n\}$ and $|\tilde D|=\tilde n$, $|\hat D|=\hat n$.
We define $\tilde \chv_j$ and $\hat \chv_j$ as the   components of $\chv_j$ 
that belong to $\tilde D$ and $\hat D$, respectively.
Under the condition that $\bfV(t)=\bfv$ and $\bfW(t)=\bfw$,
we assume that for an infinitesimal time interval of length $dt$ the evolution 
of $\bfW$ is given by the stochastic differential equation 
   \begin{equation}\label{eq:app1}
 \textstyle\bfW(t+dt)= \bfW(t) +\sum_{j=1}^m \hat\chv_j\cdot   \alpha_j(\bfv,\bfw)\cdot dt .
  \end{equation}
The evolution of $\bfV$ remains unchanged, i.e., 
$$
\pr{\bfV(t+dt)=\bfv + \tilde \chv_j\mid \bfV(t)=\bfv,\bfW(t)=\bfw}=  \alpha_j(\bfv,\bfw) \cdot dt
$$ 
The density function $h(\bfv,\bfw,t)$ of the Markov process \break
{$\{(\bfV(t),\bfW(t)),t\ge 0\}$}
can be derived in the same way as done by Horton et al.~\cite{Horton}.
Here, for simplicity we consider only the case $\hat n=1$ which means 
that $\bfw=w$ is a scalar. The  
generalization to higher dimensions is straightforward.
If $w>0$ then the following partial differential equation holds for $h$.  
$$\begin{array}{l@{\,}c@{\,}l}
&&\displaystyle \frac{\partial h(\bfv,w,t)}{\partial t} + \frac{\partial \big(h(\bfv,w,t)\cdot \sum_j \hat \chv_j\cdot \alpha_j(\bfv,w)\big)}{\partial w}\\[1.5ex]
&=&\sum\limits_j   \alpha_j(\bfv\!-\!\tilde \chv_j,w)\cdot h(\bfv \!-\!\tilde \chv_j,w,t)  
 \!-\!\sum\limits_j   \alpha_j(\bfv,w)\cdot h(\bfv,w,t).
\end{array}
$$
If $w=0$ then we have probability mass $g(\bfv,w,t)$ in state $(\bfv,w)$ where
$$\begin{array}{l@{\,}c@{\,}l}
& &\displaystyle\frac{\partial g(\bfv,w,t)}{\partial t}+\textstyle h(\bfv,w,t)\cdot \sum_j \hat\chv_j\cdot \alpha_j(\bfv,w)\\[2ex] & =&
\sum\limits_j   \alpha_j(\bfv \!-\!\tilde \chv_j,w)\cdot g(\bfv \!-\!\tilde \chv_j,w,t) 
   \!-\!\sum\limits_j   \alpha_j(\bfv,w)\cdot g(\bfv,w,t).
 \end{array}$$
As explained in-depth by Horton et al., the above equations express that probability mass must be conserved, i.e.  the change of probability 
mass in a ``cell'' with boundaries $(\bfv,w-dw)$ and $(\bfv,w+dw)$ equals the total mass of probability 
entering the cell minus  the total mass leaving the cell. 

In order to exploit the fact that the relative variance of $\bfW$ is small, we suggest an approximative solution
of the stochastic hybrid model given above. The main idea is not to compute the full density 
$h$ and the mass function $g$ but only the distribution of $V$ as well as the \emph{conditional expectations}
 $E[\bfW(t)=w\mid \bfV(t)=\bfv]$. Thus, in our numerical procedure the distribution of $\bfW$ is approximated 
by the different values $E[\bfW(t)=\bfw\mid \bfV(t)=v]$, $\bfv\in\mathbb N^{\tilde n}$ that are taken by $\bfW(t)$
with probability $\pr{\bfV(t)=\bfv}$.

Assume that at time $t$ we have  the approximation $\bfp{(t)}$ of the discrete 
stochastic model as described in Section~\ref{sec:cme}, that is,
 for all states $\bfx$ that have a probability
that is greater than $\delta$  we have $p(\bfx,t)>0$ and for all other 
states $\bfx$ we have $p(\bfx,t)=0$.
At time $t$ the expectations of one or more populations reached a
certain large population  threshold. Thus, we switch  
to a hybrid model where the large populations (index set $\hat D$) 
are represented as 
continuous deterministic variables $\bfW(t)$ while the small populations 
(index set $\tilde D$) are represented by $\bfV(t)$. We first compute 
the vector of {conditional expectations}
 $$ \textstyle
\Psi_{\bfv}(t):=E[\bfW(t)=\bfw\mid \bfV(t)=\bfv]=\sum_{\bfx:\tilde \bfx=\bfv}\bfw \cdot p(\bfx,t).$$
Here, $\tilde \bfx$ is the subvector of $\bfx$ that corresponds to $\tilde D$.
We also compute the distribution $\bfp(t)$ of $\bfV(t)$ as
$$\textstyle 
r(\bfv,t):=\sum_{\bfx:\tilde \bfx=\bfv}   p(\bfx,t).$$
Now, we integrate the system for a small time interval of length $h>0$.
This is done in three steps as described below. We will write $\Psi_{\bfv}(t')$ 
for the approximation of $E[\bfW(t')=\bfw\mid \bfV(t')=\bfv]$. 
The $i$-th element  of the $\hat n$-dimensional vector $\Psi_{\bfv}(t')$ is denoted by
$\psi_{\bfv}(i,t')$.
The value $r(\bfv,t')$ 
denotes the approximation of $\pr{\bfV(t)=\bfv}$ where $t'\in[t,t+h)$.
The vector $\bfr(t')$ contains the elements $r(\bfv,t')$.

\noindent
{\bf (1) Update distribution.} We first integrate $\bfr{(t)}$ for $h$ time units according to a CME with dimension $\tilde n$
to approximate the probabilities $\pr{\bfV(t+h)=\bfv}$  by $r(\bfv,t+h)$, that is,
$\bfr{(t+h)}$ is the solution of the system of ODEs
$$
\begin{array}{l@{\,}c@{\,}l}
 \displaystyle \frac{dr(\bfv,t')}{dt'}  
&=&\sum_j   \alpha_j\left(\bfv-\tilde \chv_j,\Psi_{\bfv-\tilde \chv_j}(t')\right)\cdot r(\bfv-\tilde \chv_j,t')\\[1.5ex]
&&-\sum_j   \alpha_j(\bfv,\Psi_{\bfv}(t'))\cdot r(\bfv,t')
\end{array}
$$
with initial condition  $\bfr{(t)}$. Note that this equation is as Eq.~\eqref{eq:CME}
except that the dimensions in $\hat D$ are removed. Moreover, the population 
sizes $\bfw$ are replaced by the conditional expectations $\Psi_{\bfv}(t')$.

\noindent
{\bf (2) Integrate.} For each state $\bfv$ with $r(\bfv,t)>\delta$, we compute an
approximation $\Phi_\bfv(t\!+\! h)$ of the conditional expectation
$$\textstyle
E[\bfW(t+h)\mid \bfV(t')=\bfv, t'\in[t,t+h)],$$
that is, we assume that the system remains in state $\bfv$ during 
$[t,t+h)$ and that the expected numbers of the large populations 
$\bfW$ change deterministically and continuously in time.
Thus, the $\hat n$-dimensional vector $\Phi_\bfv(t\!+\! h)$ is obtained by numerical integration
of the ODE 
$$\textstyle
\frac{d}{dt'} \Phi_\bfv(t') =   \sum_{j=1}^m \hat\chv_j \cdot \alpha_j(\bfv,\Phi_\bfv(t'))
$$           
with initial condition $\Phi_\bfv(t)$.
The above ODEs are similar to Eq.~\eqref{eq:exp} except that 
for $t'\in[t,t+h)$ the value
$E[\alpha_j(\bfX(t'))]$ is approximated by $\alpha_j(\bfv,\Phi_\bfv(t'))$.
For instance, if the $j$-th reaction is a bimolecular reaction that 
involves two populations with indices $i,\ell$ in $\hat D$ then  
$E[\alpha_j(\bfv,\bfW(t'))\mid \bfV(t')=\bfv]$ is
approximated by $c_j\cdot \phi_\bfv(i,t')\cdot \phi_\bfv(\ell,t')$ where 
 the two last factors are the   elements of the vector $\Phi_\bfv(t')$ corresponding 
to the $i$-th and $\ell$-th population.
Thus, in this case the correlations between the $i$-th and the $\ell$-th 
populations are not taken into account which is reasonable if the two 
populations are large. Note that the correlations \emph{are} taken into account 
when at least one population is represented as a discrete stochastic variable. If, for instance,  
$i\in\tilde D$ and $\ell\in \hat D$, then we use the approximation $c_j\cdot  v_i\cdot \phi_\bfv(\ell,t')$
 where $v_i$ is the entry in vector $\bfv$ that represents the size of the $i$-th population.

\noindent
{\bf (3) Distribute.} In order to approximate $E[\bfW(t+h)\mid \bfV(t+h)]$ by 
$\Psi_{\bfv}(t+h)$ for all states $\bfv$,
we have to replace the condition $\bfV(t')=\bfv, t'\in[t,t+h)$ by 
$\bfV(t+h)=\bfv$ in the conditional expectation $\Phi_\bfv(t\!+\! h)$ that was computed in step 2.
This is done by ``distributing'' $\Phi_\bfv(t\!+\! h)$ according to the change in the
distribution of $\bfV{(t)}$ as explained below.
The idea is to take into account that $\bfV$ enters state $\bfv$
from $\bfv'$ during the interval $[t,t+h)$. 
Assume that $[t,t+h)$ is
an infinitesimal time interval  and that $q(\bfv',\bfv,h)$, $\bfv\neq \bfv'$
 is the probability to enter $\bfv$ from $\bfv'$ within $[t,t+h)$.
Then
\begin{equation}\label{eq:condPy}
\hspace{-1ex}\begin{array}{l}
P(\bfV(t\!+\! h)\!=\! \bfv)=\sum\limits_{\bfv'\neq \bfv}   q(\bfv',\bfv,h) \cdot 
P(\bfV(t)\!=\! \bfv') \\[2ex]
\hspace{12ex} + (1-\!\!\sum\limits_{\bfv'\neq \bfv} q(\bfv,\bfv',h))\cdot P(\bfV(t)\!=\! \bfv).
\end{array} 
\end{equation}
Thus, we  approximate $E[\bfW(t+h)\mid \bfV(t+h)=\bfv]$ as
\begin{align}\label{eq:approxE}
 &\textstyle \sum\limits_{\bfv'\neq \bfv}\Phi_{\bfv'}(t\!+\! h)\cdot  q(\bfv',\bfv,h) \cdot 
P(\bfV(t)\!=\! \bfv'| \bfV(t\!+\! h)\!=\! \bfv)\\[0ex]
&\hspace{-2ex}\textstyle + \Phi_\bfv(t\!+\! h)\cdot (1\!-\!\!\!\sum\limits_{\bfv'\neq \bfv} q(\bfv,\bfv',h))
 \cdot P(\bfV(t)\!=\! \bfv| \bfV(t\!+\! h)\!=\! \bfv).\notag
 \end{align}
%
Obviously,
  we can make use of the current approximations $\bfr{(t)}$ 
and $\bfr{(t+h)}$ to compute
the conditional probabilities $P(\bfV(t)\!=\! \bfv'| \bfV(t+ h)\!=\! \bfv)$.
For a small time step $h$,
$q(\bfv',\bfv,h)\approx h\cdot \alpha_j(\bfv',\Psi_{\bfv'}(t))$                                                               
if
$\bfv'=\bfv+\tilde\chv_j$ and $q(\bfv',\bfv,h)\approx 0$ otherwise.
 
Using Eq.~\eqref{eq:approxE}, we compute the approximation
$\Psi_{\bfv}(t+h)\approx E[\bfW(t\!+\! h)| \bfV(t\!+\! h)=\bfv]$  as
\begin{equation}\label{eq:distribute}
\hspace{-0.5ex}\begin{array}{l}
\Psi_{\bfv}(t+h)=  \\[1ex] 
\hspace{2ex}\sum_{j} \Phi_{\bfv\!-\! \tilde\chv_j}\!(t\!+\! h)\cdot \frac{p(\bfv\!-\! \tilde\chv_j,{t})}{p(\bfv,t+h)} 
\cdot
 \alpha_j(\bfv\!-\! \tilde\chv_j,\Psi_{\bfv-\tilde\chv_j}(t)) \cdot h
 \\[1.5ex]
\hspace{2ex}+\Phi_{\bfv}(t\!+\! h)\cdot \frac{p(\bfv,{t})}{p(\bfv,t+h)}  (1-\sum_j\alpha_j(\bfv,\Psi_{\bfv}(t)) \cdot h   ).
\end{array}\hspace{-2ex}
\end{equation}
Note that the first sum runs over all direct predecessors ${\bfv\!-\! \tilde\chv_j}$ of
$\bfv$.   
 \begin{figure}[t]
\begin{center}
 	 \includegraphics[width=4cm]{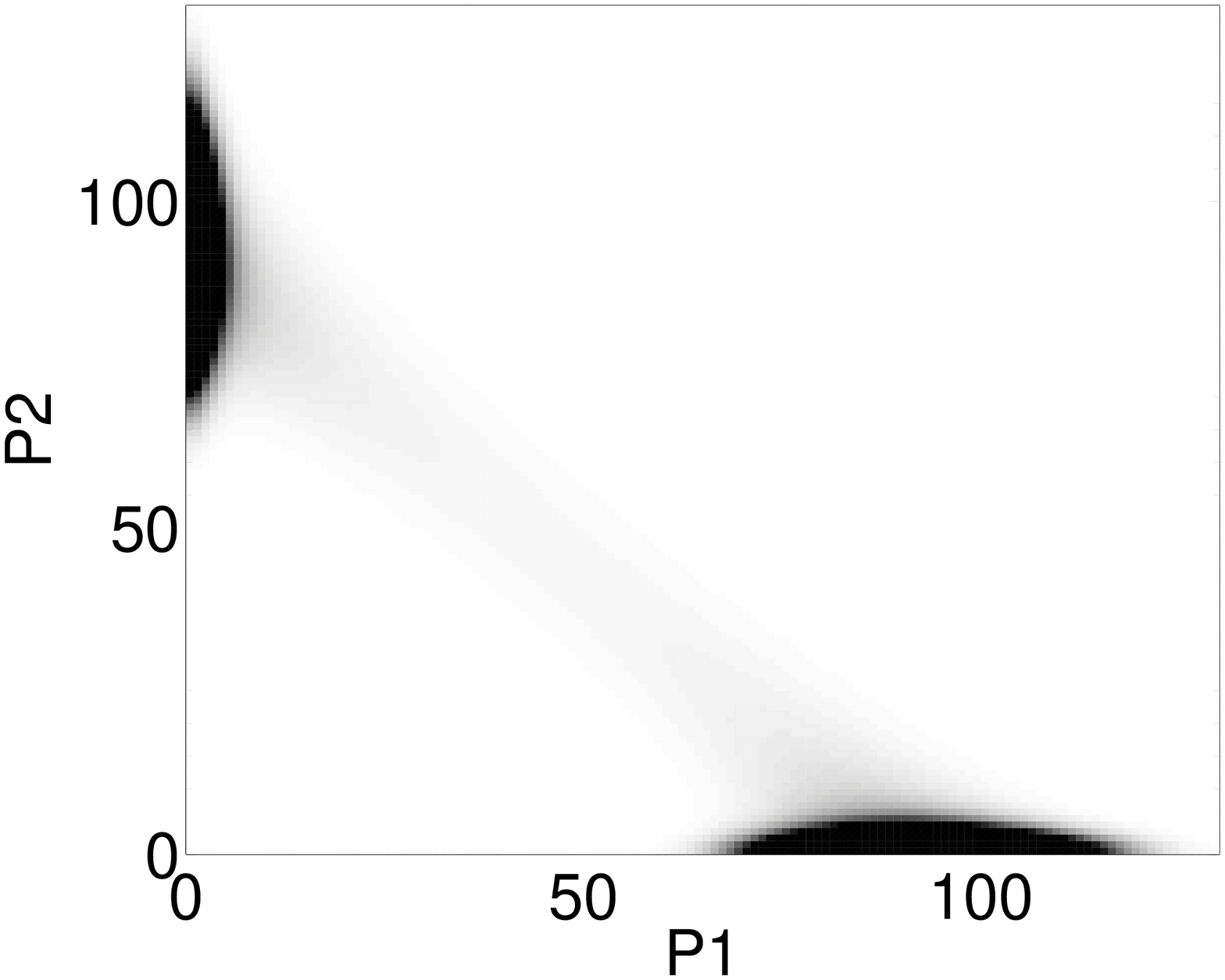} 
\includegraphics[width=4cm]{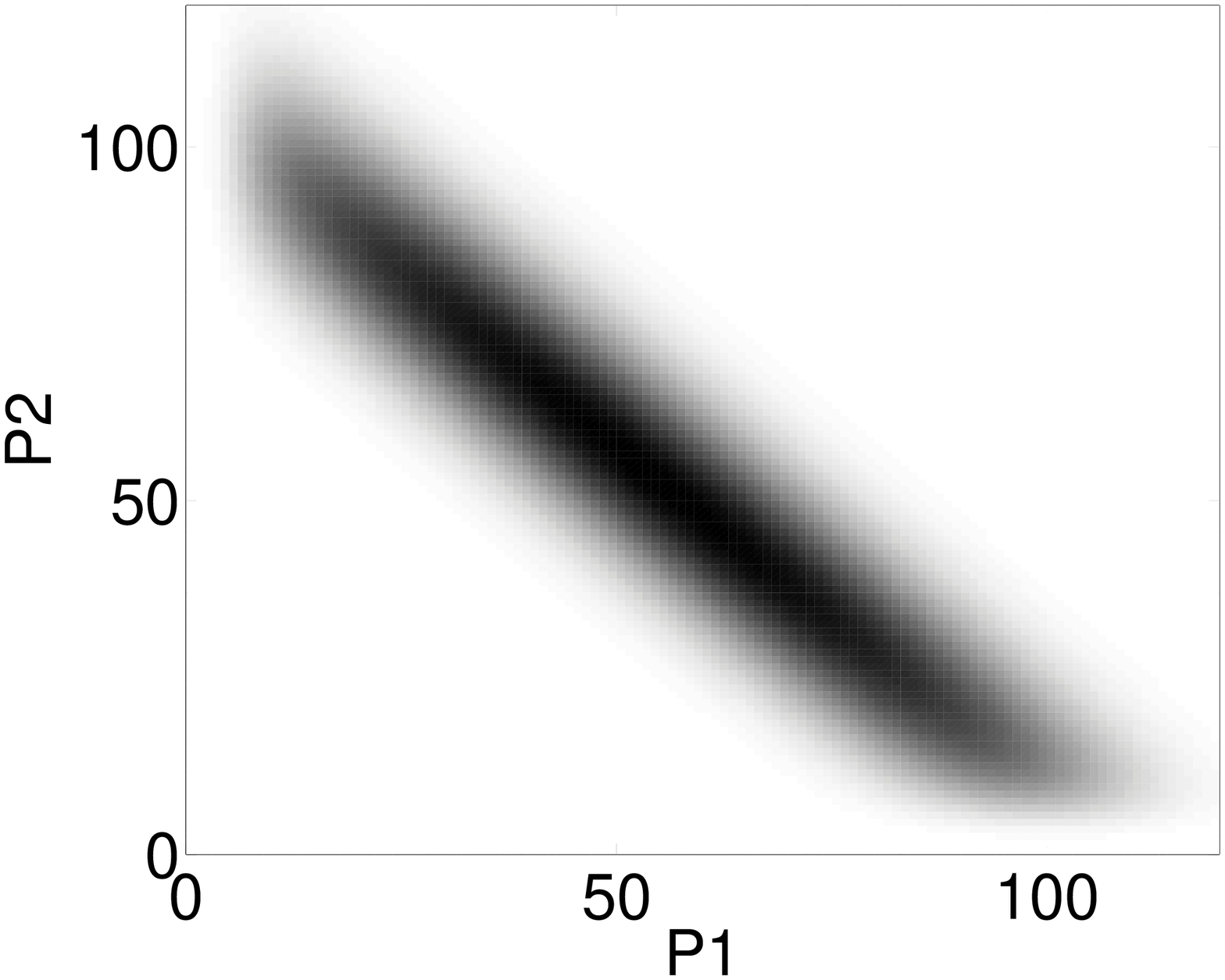}\vspace{-3ex}
\end{center}		\caption{Probability distribution of the exclusive switch in Ex.~1. 
\label{fig:switchplot}} \vspace{-2ex}
 \end{figure}
\begin{examplecont}
In the exclusive switch the expected number of molecules of type $P_1$ and/or $P_2$ 
may become high, depending on the chosen parameters. If, for instance, $c_1\!=\!c_2\!=\!c_9\!=\!c_{10}\!=\!0.5$, 
$c_3\!=\!c_4\!=\!c_7\!=\!c_8\!=\!0.005$,   $c_5\!=\!c_6\!=\!0.01$, and we start initially without any proteins, i.e. with probability one
in state $\bfy=(0,0,1,0,0)$, then after 500 time units most of the probability mass is located around the states 
$\bfx\!=\!(92,2,0,1,0)$   and $\bfx\!=\!(2,92,0,0,1)$ (compare the plot in Fig.~\ref{fig:switchplot}, left).  
Note that $x_3\!=\!0, x_4\!=\!1, x_5\!=\!0$ refers to the case that a molecule of type $P_1$ is bound to the promotor 
and $x_3\!=\!x_4\!=\!0, x_5\!=\!1$ refers to the case that a molecule of type $P_2$ is bound to the promotor. 
Since for these parameters  the system is symmetric, the expected populations of  $P_1$ and $P_2$ are 
identical. Assume that at a certain time instant, both populations reach the threshold from which on we approximate them 
by continuous deterministic variables (we consider the unsymmetric case later, in Section~\ref{sec:casestudy}).
 The remaining discrete model then becomes finite since only $P_1$ and $P_2$ have an infinite range in the original 
model ($\hat n=2, \tilde n=3$).
More precisely, it contains only 3 states, namely the state $\bfv_1$ where the promotor is free ($x_3\!=\!1, x_4\!=\!x_5=0$),
the  state $\bfv_2$ where $P_1$ is bound to the promotor ($x_3\!=\!0, x_4\!=\!1, x_5\!=\!0$), and the state $\bfv_3$ 
where $P_2$ is bound to the promotor ($x_3\!=\!x_4\!=\!0, x_5\!=\!1$), see also Fig.~\ref{fig:switch}. 
For $i\in\{1,2\}$ let $W_i(t)$ be the population size of $P_i$.
The differential equations which are used to approximate 
the conditional expectations $\Psi_{\bfv_j}(t+h)$, $j\in\{1,2,3\}$ are 
$$\begin{array}{lcl}
\frac{d \phi_{\bfv_1}(i,t') }{dt'}&= &c_i- c_{2+i}\cdot \phi_{\bfv_1}(i,t')-c_{4+i}\cdot \phi_{\bfv_1}(i,t')\\[0.5ex]
\frac{d \phi_{\bfv_2}(i,t') }{dt'}&= &- c_{2+i}\cdot \phi_{\bfv_2}(i,t')+(c_{7}+c_{9})\cdot (2-i) 
\\[0.5ex]
\frac{d \phi_{\bfv_3}(i,t') }{dt'}&= &- c_{2+i}\cdot \phi_{\bfv_3}(i,t')+
(c_{8}+c_{10})\cdot (i-1+) 
\end{array}
$$
where $\phi_{\bfv}(1,t')$ and $\phi_{\bfv}(2,t')$ are the elements of the vector $\Phi_{\bfv_1}(t')$. 
 Note that each of the 3 states $\bfv_1,\bfv_2,\bfv_3$ has a system of 
two differential equations,
one for $P_1$ and one for $P_2$. The transition rates in the 
discrete stochastic part of the model are illustrated in Fig.~\ref{fig:vs}.
 \begin{figure}[t]
\begin{center}
 \begin{picture}(80,30)
  \gasset{Nw=6,Nh=5,Nmr=5}
\node(p1)(5,12){$\bfv_1$}
\node(p2)(55,12){$\bfv_2$}
 \node(p3)(75,12){$\bfv_3$}
\drawedge[curvedepth=2](p1,p2){$c_5\cdot \psi_{\bfv_1}(1,t)$}
 \drawedge[curvedepth=2](p2,p1){$c_7$}
\drawedge[ELside=l,curvedepth=12](p1,p3){$c_6\cdot \psi_{\bfv_1}(2,t)$}
\drawedge[ELside=l,curvedepth=8](p3,p1){$c_8$}
 
 \end{picture}
\vspace{-5ex}
\end{center}		\caption{The discrete stochastic part 
of the stochastic hybrid model of Ex.~1.\label{fig:vs}}
\vspace{-3ex}
\end{figure}
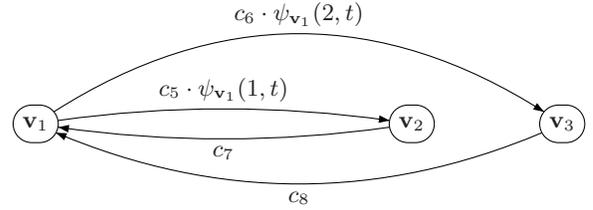
\begin{table*}[t]
\renewcommand{\arraystretch}{1.1}
 \begin{center}
\begin{tabular}{|r|r|r|r|r|r|r|r|r|r|r|r|} 
 \hline
pset& \multicolumn{3}{c|}{purely stochastic} & 
 \multicolumn{6}{c|}{stochastic hybrid}& \multicolumn{2}{c|}{purely determ.} \\
\cline{2-12}
 & ex. time & $|\sig|$ &error & pop. thres. & ex. time & $|\sig|$ & m1&m2&m3 & ex. time & m1\\
     \hline
1a& 11h 46min  & $8 \cdot 10^5$ & $7 \cdot 10^{-5}$ &  50 & 15sec  & $4 \cdot 10^2$ & 0.005 & 0.2 & 0.30 & 1sec & 0.03 \\
\cline{5-10}
&           &                &                   & 100 & 1min 50sec & $3 \cdot 10^3$ & 0.004 & 0.2 & 0.30 & & \\
\hline
1b& 7min 43sec & $5 \cdot 10^4$ & $7 \cdot 10^{-7}$ &  50 & 1min 19sec  & $6 \cdot 10^3$ & 0.01 & 0.19 & 0.30 & 1sec & 0.03 \\
\cline{5-10}
&        &                &                   & 100 & 2min 50sec & $3 \cdot 10^4$ & 0.01 & 0.19 & 0.30 &   &  \\
\hline
2& 4h 51min & $2 \cdot 10^5$ & $4 \cdot 10^{-5}$ &  50 & 25sec & $4 \cdot 10^2$ & 0.06 & 0.08 & 0.09 & 1sec & 0.45 \\
\cline{5-10}
&         &                &                   & 100 & 28sec & $6 \cdot 10^2$ & 0.06 & 0.07 & 0.09 &  &  \\
\hline
3&  2min 21sec  & $7 \cdot 10^5$ & $6 \cdot 10^{-5}$ &  50 &  18sec & $6 \cdot 10^3$ & 0.02 & 0.08 & 0.16 & 1sec & 0.05 \\
\cline{5-10}
&         &                &                   & 100 & 1min 41sec & $4 \cdot 10^4$ & 0.01 & 0.05 & 0.12 & & \\
\hline
  \end{tabular}              \end{center}\vspace{-3ex}
   \caption{Results for the exclusive switch example.\label{tab:exswitch}}
   \end{table*}
Thus, after solving the differential equations above to compute 
$\Phi_{\bfv_j}(t+h)$ we obtain the vector $\Psi_{\bfv_j}(t+h)$ of the two conditional 
expectations for $P_1$ and $P_2$
from distributing $\Phi_{\bfv_1}(t+h)$, $\Phi_{\bfv_2}(t+h)$, $\Phi_{\bfv_3}(t+h)$
among the 3 states as  defined in Eq.~\eqref{eq:distribute}.
For the parameters used in Fig.~\ref{fig:switch}, left,
the conditional expectations of the states $\bfv_2$ and $\bfv_3$  accurately predict
the two stable regions where most of the probability mass is located. 
The state $\bfv_1$ has small probability and its conditional expectation is located
 between the two stable regions. It is important to point out that, for this example, a purely deterministic 
solution cannot detect the bistability because the deterministic model has a single 
steady-state~\cite{exswitch}.
Finally, we remark that in this example the number of states in the reduced discrete model is 
very small. If, however, populations with an infinite range but small expectations 
are present, we use the truncation described in Section~\ref{sec:cme} to keep 
the number of states small.
 \end{examplecont}   
   
 If at time $t$ a population, say the $i$-th population,
 is represented by its conditional expectations, it is 
possible to switch back to the original discrete stochastic treatment.
This is done by  adding an entry to the states $\bfv$ for the $i$-th dimension.
This entry then equals $\psi_{\bfv}(i,t)$. 
This means that at this point we assume that the conditional probability 
distribution has mass one for the value $\psi_{\bfv}(i,t)$.
 Note that here switching back and forth  between discrete stochastic 
and continuous deterministic representations is based on a population 
threshold. Thus, if the expectation of a population oscillates we may switch 
back and forth in each period.

 \section{Experimental Results}\label{sec:casestudy}
 We implemented the numerical solution of the stochastic hybrid
model described above in C++ as well as the fast solution of the 
discrete stochastic model described in Section~\ref{sec:cme}.
In our implementation we dynamically switch the representation of a random 
variable whenever it reaches a certain population threshold. 
 We ran experiments with two different thresholds (50 and 100) 
on an Intel 2.5GHz   Linux workstation with 8GB of RAM. 
 In this section we present 3 examples to that we applied 
 our algorithm, namely the exclusive switch, Goutsias' model, and 
a predator-prey model. Our most complex example   
has 6 different chemical species and 10 reactions.
We compare our results to  a purely stochastic solution where switching is 
turned off as well as to a purely deterministic solution. 
 For all experiments, we fixed the cutting threshold 
$\delta=10^{-14}$ to truncate 
the infinite state space as explained in Sec.~\ref{sec:cme}.


%
  
\begin{table*}[t]
\renewcommand{\arraystretch}{1}
 \begin{center}
\begin{tabular}{|r|r|r|r|r|r|r|r|r|r|r|r|} 
 \hline
model& \multicolumn{3}{c|}{purely stochastic} & 
 \multicolumn{6}{c|}{stochastic hybrid}& \multicolumn{2}{c|}{\hspace{-2ex}\begin{tabular}{c}purely\\ determ.\end{tabular}
\hspace{-2ex} } \\
\cline{2-12}
& ex. time & $|\sig|$ &error & \hspace{-2ex}\begin{tabular}{c}pop.\\ thres.\end{tabular}
\hspace{-2ex} & ex. time & $|\sig|$ & m1&m2&m3&  \hspace{-2ex}\begin{tabular}{c}ex.\\ time\end{tabular}
\hspace{-2ex}  & m1\\
     \hline
\hspace{-1ex}Goutsias\hspace{-1ex}& 1h 16min & $1 \cdot 10^6$ & $4 \cdot 10^{-7}$ &  50 & 8min 47sec & $1 \cdot 10^5$ &  0.001 &  0.07 &  0.13 & 1sec & 0.95 \\
\cline{5-10}
&         &                &                   & 100 & 48min 57sec & $6 \cdot 10^5$ & 0.0001 & 0.0003 & 0.001   & & \\
\hline
p.-prey& 6h 6min & $5 \cdot 10^5$  & $1 \cdot 10^{-7}$  & 50  &  8min 56sec  & $2 \cdot 10^4$ &  0.06 &  0.15  & 0.27  & 1sec &  0.86 \\
\cline{5-10}
&            &                &                   &   100   & 1h 2min  &  $8 \cdot 10^4$ &  0.04 &  0.11 & 0.23 & & \\
\hline
  \end{tabular}              \end{center}\vspace{-3ex}
   \caption{Results  for Goutsias' model and the predator-prey model.\label{tab:goutsias}}
   \end{table*}
{\bf Exclusive Switch.}
We chose different parameters for the exclusive switch in order
to test whether our hybrid approach works well if\vspace{-1ex}
\begin{compactitem}
 \item[1)] the populations of $P_1$ and $P_2$ are large (a) or  small (b),
\item[2)] the model is unsymmetric (e.g. $P_1$ is produced at a  higher rate than $P_2$ and degrades at a  slower rate than $P_2$),
\item[3)] the bistable form of the distribution is destroyed (i.e. promotor binding   is less likely, unbinding is more likely).
\end{compactitem} 
The following table lists the  parameter sets (psets):\vspace{-2ex}
\begin{center}
$
\begin{array}{@{\vrule\,}c@{\,\vrule\,}c@{\,\vrule\,}c@{\,\vrule\,}c@{\,\vrule\,}c@{\,\vrule\,}c@{\,\vrule\,}c@{\,\vrule\,}c@{\,\vrule\,}c@{\,\vrule\,}c@{\,\vrule\,}c@{\,\vrule}}
\hline
 \mbox{pset} & c_1 & c_2 & c_3 &c_4 & c_5 & c_6 & c_7&c_8&c_9&c_{10}\\
\hline
1a & 5 & 5 &0.0005 & 0.0005 & 0.1 & 0.1& 0.005&0.005 & 5&5\\
\hline
1b & 0.5 & 0.5 &0.0005 & 0.0005 & 0.1 & 0.1& 0.005&0.005 & 0.5&0.5\\
\hline
2 & 5 & 0.5 &0.0005 & 0.005 & 0.1 & 0.1& 0.005&0.005 & 5&0.5\\
\hline
3 & 0.5 & 0.5 &0.0005 & 0.0005 & 0.01 & 0.01& 0.1&0.1 & 0.5&0.5\\
\hline
\end{array}
$\end{center}\vspace{-2ex}
We chose a  time horizon of $t=500$ for all parameter sets.
Note that in the case of pset 3 the probability distribution forms a thick line in the state space
(compare the plot in Fig.~\ref{fig:switchplot}, right).   
We list   our results in Table~\ref{tab:exswitch} 
where the first column refers to the  
parameter set. Column 2 to 4 list the results of a purely stochastic solution (see 
Section~\ref{sec:cme}) where ``ex. time'' refers to the execution time, $|\sig|$ to the average size of the 
set of significant states and ``error'' refers to the amount of probability mass lost due to the 
truncation with threshold $\delta$, i.e. $1-\sum_{\bfx\in\sig}p(\bfx,t)$.
 The columns 6-10 list the results of our stochastic hybrid approach
and column 5 lists
the population threshold used for switching in the representations 
in the stochastic hybrid model.
Here, ``m1'', ``m2'', ``m3'' refer to the relative error of the first three moments of the joint probability 
distribution at the final time instant. 
For this, we compare the (approximate) solution of the hybrid model with the 
solution of the purely stochastic model.  
Since we have five species, we simply take the average relative error over all species.
Note that even if a species is represented by its conditional expectations, we can
approximate its first three moments by
$$
\textstyle E[\bfW(t)^i]\approx\sum_{\bfv} \left(\Psi_{\bfv}(t)\right)^i\cdot r(\bfv,t)
$$
where the $i$-th power of the vectors are taken component-wise.
Finally, in the last two columns we list the results of a purely deterministic solution as explained in 
Section~\ref{sec:ODE}. The last column refers to the average relative error of the expected populations
when we compare the purely deterministic solution to the purely stochastic solution.
Note that the deterministic solution of the exclusive switch yields an accurate approximation of the
first moment (except for pset 2) because of the symmetry of the model. 
It does, however, not reveal the bistability of the distribution.
As opposed to that, the hybrid solution \emph{does} show this important property. 
For pset 1 and 3, the 
conditional expectations of the 3 discrete states 
are such that two of them match exactly the 
two stable regions where most of the probability mass is located 
(see also Example 1 in Sec.~\ref{sec:stochhybrid})  . 
The remaining conditional expectation
of the state where the promotor region is free has small probability and predicts a conditional 
expectation between the two stable regions.
The execution time of the purely stochastic  approach is high in the case of pset 1a,
because the expected populations of $P_1$ and $P_2$ are high. This yields large 
sizes of $\sig$ while we iterate over time. During the hybrid solution, we switch when the 
populations reach the threshold and the size of $\sig$ drops to 3. Thus, the average 
  number of significant states is much smaller. In the case of pset 1b, the expected 
populations are small and we use a deterministic representation for protein populations 
only during  a short time interval (at the end of the time horizon).  
For   pset 2, the accuracy of the purely deterministic solution is poor because the model 
is no longer symmetric. 
The accuracy of the hybrid solution on the other hand is independent of the symmetry of the model.

{\bf Goutsias' Model.}
  In~\cite{goutsias}, Goutsias defines a model
for the transcription regulation of a
repressor protein in bacteriophage $\lambda$.
This protein is responsible for maintaining lysogeny
of the $\lambda$ virus in E. coli.
The model involves 6 different species and the following 10 reactions.
\begin{center}\vspace{-4ex}
$ \begin{array}{r@{: \  }l@{\,}c@{\,}l@{\hspace{2ex}}r@{: \  }l@{\,}c@{\,}l}
1 &  \mbox{\small RNA} & \to & \mbox{\small RNA} \! +\! \mbox{\small M} &
6 & \mbox{\small DNA.D} & \to & \mbox{\small DNA} \! +\!  \mbox{\small D}\\
2 &  \mbox{\small M}  & \to & \emptyset &  7 &  \mbox{\small DNA.D} \! +\!  \mbox{\small D} & \to & \mbox{\small DNA.2D}\\
3 &  \mbox{\small DNA.D}  & \to & \mbox{\small RNA}  \! +\! \mbox{\small DNA.D} &8 &  \mbox{\small DNA.2D} & \to & \mbox{\small DNA.D}  \! +\! \mbox{\small 
D}\\
4 &  \mbox{\small RNA}  & \to &  \emptyset &9 &  \mbox{\small M} + \mbox{\small M} & \to & \mbox{\small D}\\
 5 &  \mbox{\small DNA}  \! +\! \mbox{\small D} & \to & \mbox{\small DNA.D}
&
10 &  \mbox{\small D} & \to & \mbox{\small M} \! +\! \mbox{\small M}\\
  \end{array}
  $\vspace{-4ex}
\end{center}
 We used the following parameters  
 that     differ from the original parameters  
  used in~\cite{goutsias}
 in that they increases the number of 
 RNA molecules (because with the original parameters, all populations remain small).  
\begin{center}\vspace{-2ex}
$
\begin{array}{@{\vrule\,}c@{\,\vrule\,}c@{\,\vrule\,}c@{\,\vrule\,}c@{\,\vrule\,}c@{\,\vrule\,}c@{\,\vrule\,}c@{\,\vrule\,}c@{\,\vrule\,}c@{\,\vrule\,}c@{\,\vrule}}
\hline
  c_1 & c_2 & c_3 &c_4 & c_5 & c_6 & c_7&c_8&c_9&c_{10}\\
\hline
  0.043 & 7\mbox{e{-4}} & 71.5 & \mbox{3.9e{-6}} & 0.02 & 0.48& \mbox{2e{-4}}
&\mbox{9e{-12}} & 0.08&0.5\\ 
\hline
\end{array}
$\end{center}\vspace{-2ex}

  \begin{figure}[t]
\begin{center}
 \vspace{-1ex}\hspace{-2ex}	 \includegraphics[width=4.2cm]{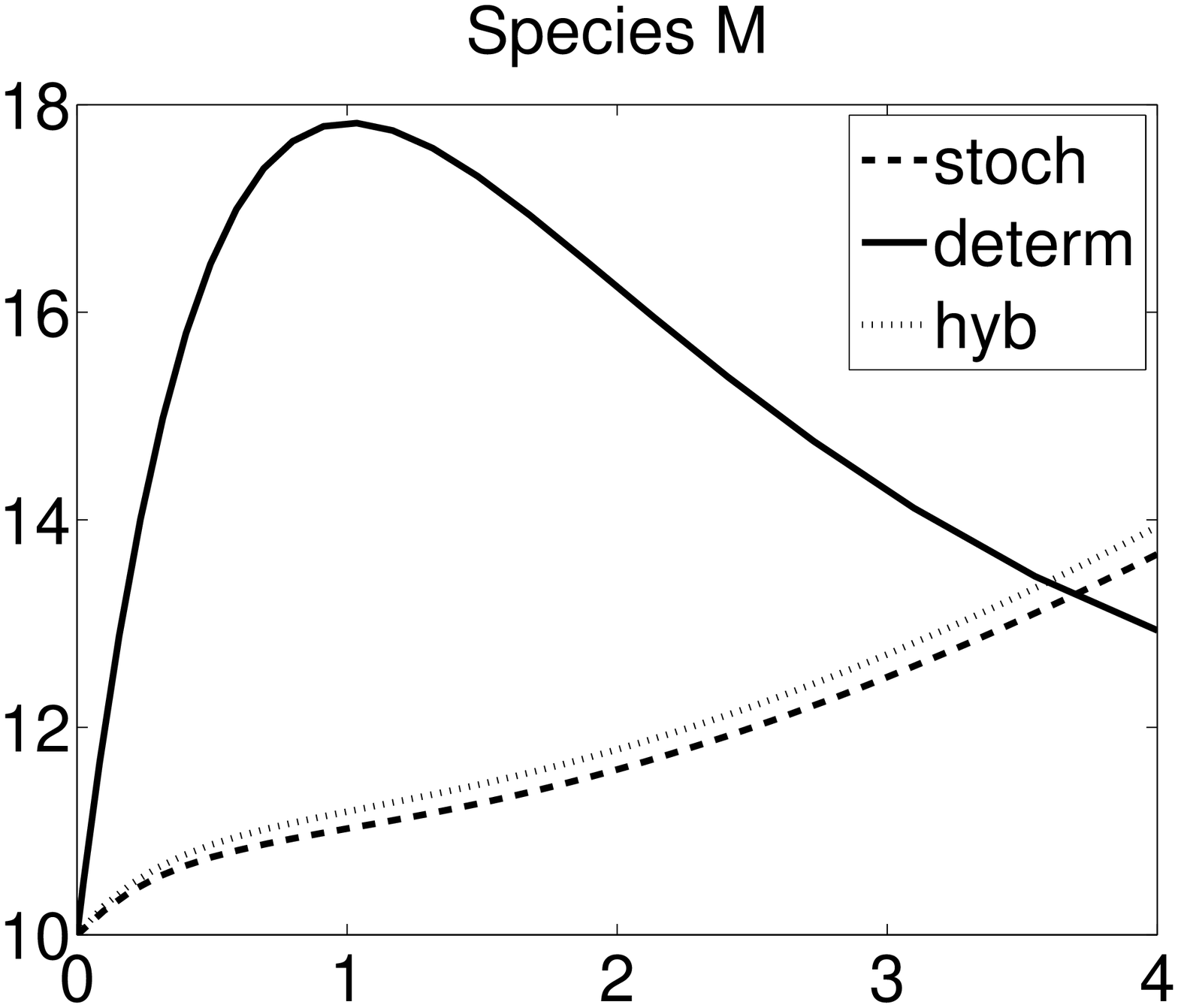} \hspace{-2ex}
 \includegraphics[width=4.2cm]{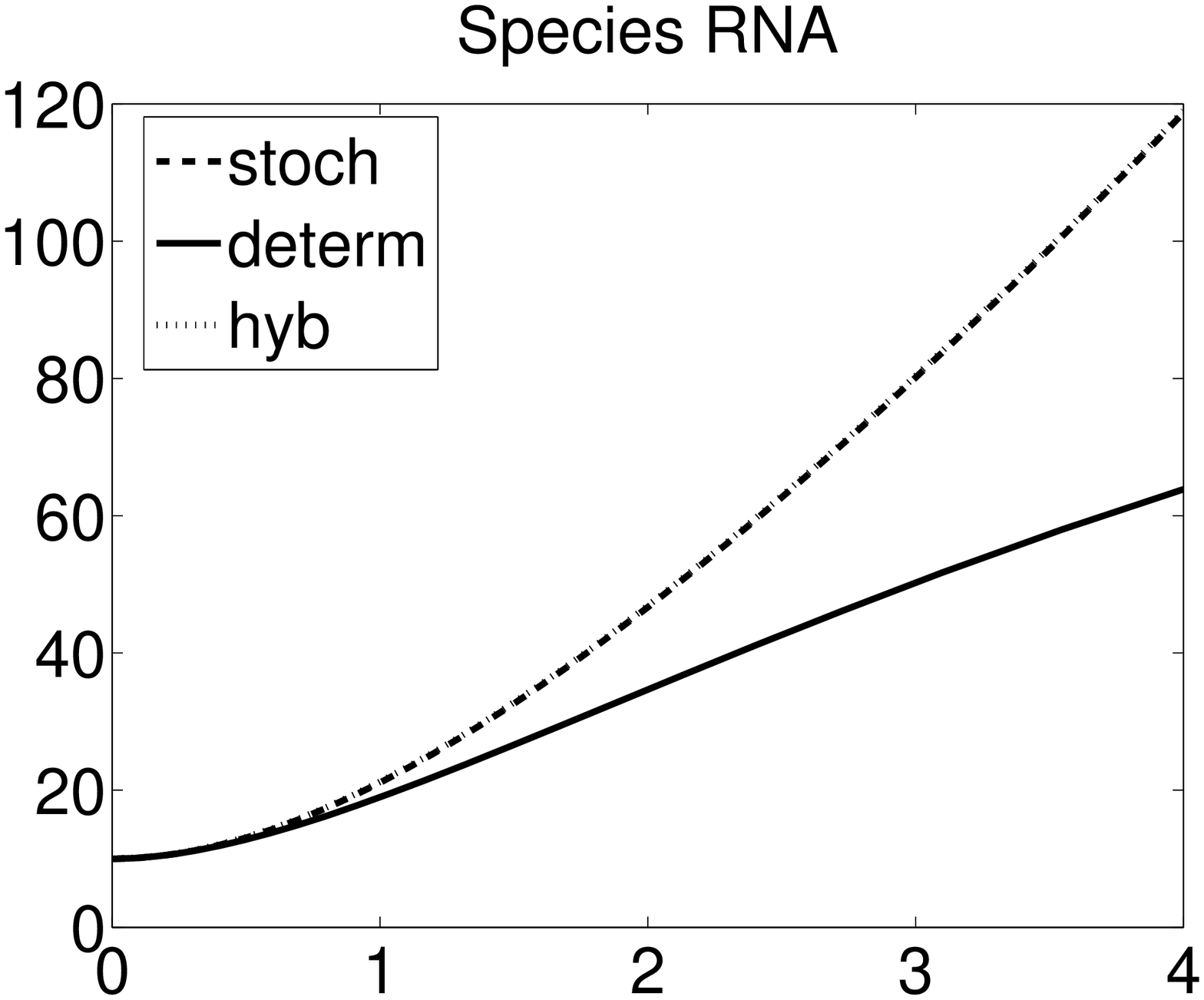} \hspace{-2ex}\vspace{-3ex}
   \end{center}		\caption{Expected populations in Goutsias' model.\label{fig:goutsias}} \vspace{-3ex}
 \end{figure}

Table~\ref{tab:goutsias} shows the results for the 
Goutsias' model where we use the same column labels as above.
 We always start initially with 10 molecules of RNA, M, and D, as well as 
2 DNA molecules. We choose the time 
horizon  as  $t=4$.
Note that the hybrid solution as well as the purely deterministic solution 
are feasible for much longer time horizons. 
The increase of the size of the set of significant states 
makes the purely stochastic solution infeasible for longer
time horizons. As opposed to that the memory requirements of the hybrid solution  
remain tractable. In Fig.~\ref{fig:goutsias} we plot the means of two of the six
species obtained from the purely stochastic (stoch), purely deterministic (determ),
and the hybrid (hyb) solution.
Note that a purely deterministic solution yields very poor accuracy 
(relative error  of the means is 95\%).

 {\bf Predator Prey.}
 We apply our algorithm to the predator prey model  described in~\cite{gillespie77}.
 It involves two species $A$ and $B$ and the reactions are
 $A \to 2 A$, $A+B\to 2 B$, and $B\to \emptyset$.
 The model shows sustainable periodic oscillations until 
 eventually the population of $B$ reaches zero. 
We use this example to test the switching mechanism of our
algorithm. 
 We choose rate constants $c_1=1$, $c_2=0.03$,  $c_3=1$ and
 start initially with 30 molecules of type $A$ and 120 molecules of type $B$.
 For a population threshold of 50, we start with a stochastic representation of $A$ and 
 a deterministic representation of $B$. Then, around  time $1.3$ 
we switch to a purely stochastic representation since the expectation of $B$ becomes less than 50. 
Around time $t=6.1$ we switch the representation of 
$A$ because $E[A(t)] > 50$, etc.
   We present our detailed results in Table~\ref{tab:goutsias}.
Similar to Goutsias' model, the deterministic solution has a 
high relative error whereas the hybrid solution yields accurate results.
 \section{Conclusion}
 We presented a stochastic hybrid model for the analysis of networks of chemical reactions.
This model is based on a dynamic switching between a  discrete stochastic  
and a continuous deterministic representation of the chemical populations. 
Instead of solving the underlying partial differential equation, we propose a 
fast numerical procedure that exploits 
the fact that for large populations the conditional expectations give appropriate 
approximations.  Our experimental results substantiate the usefulness of the method.
As future work we plan to include a diffusion approximation for populations of intermediate size.


%
\bibliographystyle{abbrv}
\bibliography{bio}  
 
 

\balancecolumns
\end{document}